\documentstyle[12pt]{article}

\newcommand{\be}{\begin{equation}}
\newcommand{\ee}{\end{equation}}
\def\aprle{\buildrel < \over {_{\sim}}}
\def\aprge{\buildrel > \over {_{\sim}}}
\begin{document}
\topmargin 0pt
\oddsidemargin=-0.4truecm
\evensidemargin=-0.4truecm      
\renewcommand{\thefootnote}{\fnsymbol{footnote}}     
\setcounter{page}{0}
\begin{titlepage}     
\begin{flushright}
SISSA 27/97/A-EP \\
IC/97/16\\
\end{flushright}
\vspace{0.6cm}
\begin{center}
{\Large \bf Resonant spin--flavour precession of neutrinos and pulsar 
velocities}\\ \vspace{0.9cm}
{\large E.Kh. Akhmedov  
\footnote{On leave from NRC ``Kurchatov Institute'', Moscow 123182, Russia. 
e-mail address: akhmedov@sissa.it}}\\ 
\vspace{0.2cm}
{\em International Centre for Theoretical Physics \\
 Strada Costiera 11, I-34100 Trieste, Italy}\\ 
\vskip .5cm
{\large A. Lanza
\footnote{e-mail address: lanza@sissa.it} and D.W. Sciama 
\footnote{Also at International Centre for 
Theoretical Physics, Strada Costiera 11, I-34100 Trieste, Italy. 
e-mail address: sciama@sissa.it}}\\
\vspace{0.2cm}
{\em Scuola Internazionale Superiore di Studi Avanzati\\
Via Beirut 2--4, I-34014 Trieste, Italy} \\
\end{center}
\vspace{0.4cm} 
\begin{abstract}
Young pulsars are known to exhibit large space velocities, up to $10^3$  
km/s.  We propose a new mechanism for the generation of these large  
velocities based on an asymmetric emission of neutrinos during the 
supernova explosion. The mechanism involves the resonant spin--flavour 
precession of neutrinos with a transition magnetic moment in the magnetic 
field of the supernova. The asymmetric emission of neutrinos is due 
to the distortion of the resonance surface by matter polarisation effects 
in the supernova magnetic field. 
The requisite values of the field strengths and neutrino parameters are 
estimated for various neutrino conversions caused by their Dirac or 
Majorana--type transition magnetic moments. 
\end{abstract}
\centerline{PACS number(s): 97.60.Gb, 14.60.Pq, 14.60.St}
\end{titlepage}
\renewcommand{\thefootnote}{\arabic{footnote}}
\setcounter{footnote}{0}
\newpage
\section{Introduction}  
In this paper we propose a new mechanism for generating large birth 
velocities of pulsars which is related to the possible existence 
of transition magnetic moments of neutrinos. 
Observations imply   
that pulsars have rapid proper motions with mean space velocity 
of $450\pm 90$ km/s \cite{Lyne}. In particular, observations in young 
supernova remnants have identified pulsars with velocities up to 
$900$ km/s \cite{CF}. 
Such high velocities of young pulsars are most probably associated 
with the supernova event in which the pulsar is born.  
Many different models have been formulated to explain the origin of these 
velocities: asymmetric collapse \cite{Collapse}, asymmetry due to 
misalignment of dipole magnetic field and rotation axis \cite{Harrison},  
a ``runaway star'' produced by a supernova explosion in a close binary star 
\cite{Gott}, and recoil momentum from the asymmetric production of neutrinos 
\cite{Field-weak} in the supernova's magnetic field. The latter possibility 
is especially interesting since neutrinos carry away more than $99\%$ of the 
supernova's gravitational binding energy and so even a $\sim 1\%$ asymmetry 
in the neutrino emission could lead to the observed recoil 
velocities of pulsars. However, this mechanism requires a magnetic field 
$B\aprge 10^{16}$ G which may be somewhat too strong for supernovae. 
It is also possible that the asymmetry in the neutrino momenta will be 
washed out by multiple scattering, absorption and re-emission of the 
neutrinos in the core of the supernova \cite{Vol}.  

Recently, a very interesting new mechanism for the generation of the 
pulsar velocities has been proposed by Kusenko and Segr\`e (KS) \cite{KS1}. 
The idea is that the neutrino emission from a cooling proto--neutron star can 
be asymmetric due to resonant neutrino oscillations (MSW effect \cite{MSW}) 
in the supernova's magnetic field. The anisotropy of the neutrino emission is 
driven by the polarization of the medium in a strong magnetic field, which 
distorts the resonance surface.  
According to KS, this mechanism requires magnetic fields $B\aprge 3\times 
10^{14}$ G. 
However, this estimate 
was recently 
criticised by Qian who showed that in fact magnetic fields $B\aprge 10^{16}$ 
G are necessary \cite{Qian}. We will come back to this question later on.  
An advantage of the KS scenario is that it is free from the 
above--mentioned shortcoming of the asymmetric production mechanism: the 
neutrinos carrying the momentum asymmetry are free--streaming from the 
supernova and therefore the asymmetry is hardly affected by the interaction 
of neutrinos with matter. 
This mechanism is very attractive since it predicts a well defined 
correlation between the strength of the magnetic field and the observed 
space velocites of pulsars \cite{KS2}. 
As a follow-up, in ref. \cite{KS3} the same authors have considered
the effects of sterile-to-active neutrino oscillations, where now 
neutral--current effects are important. 
Thus, as was stressed in \cite{KS1,KS2,KS3}, pulsar motions may be a  
valuable source of information on neutrino properties. One should therefore 
study all possible mechanisms of asymmetric neutrino conversions in 
supernovae that might be the cause of the observed velocities of pulsars. 

The mechanism that we propose here is similar 
to the KS one -- it is based on the observation that matter polarisation 
in the strong magnetic field of a supernova distorts the resonance surface 
of the neutrino conversion. The difference from the KS scenario is in the 
nature of the neutrino transition involved: in our case it is the resonant  
spin--flavour precession of neutrinos due to their transition magnetic 
moments \cite{Akhm1,LM} rather than neutrino oscillations\footnote{The 
resonantly enhanced spin--flavour precession of neutrinos in supernovae 
has been considered in the literature \cite{LM,RSFP}, but no implications 
for pulsar velocities were discussed there.}.
We show here that our proposed mechanism can account for the observed large 
space velocities of pulsars provided that the neutrino transition magnetic 
moment satisfies $\mu_\nu\aprge (10^{-15}-10^{-14})\mu_B$, the neutrino 
mass is less than or about 16 keV and the supernova magnetic magnetic field  
$B\aprge 4\times 10^{15}$ G. Notice that the required magnetic field is 
slightly weaker than the one necessary for the KS mechanism \cite{Qian}.  
Such large fields are presently considered possible in supernovae \cite{DT}.  

If the supernova magnetic field has a noticeable ``up--down'' asymmetry,  
neutrino spin or spin--flavour precession can 
affect differently neutrinos emitted in the upper and lower hemispheres. 
As was pointed out by Voloshin \cite{Vol}, this could also be the reason 
for the observed birth velocities of pulsars. In this paper we restrict 
ourselves to the case of symmetric magnetic fields; 
possible effects of neutrino conversions in asymmetric magnetic fields will 
be considered elsewhere \cite{ALS2}. 

\section{Neutrino potentials in polarised media}
In a medium containing a magnetic field the particles of matter have in 
general nonzero average spin. This spin polarisation contributes to the 
neutrino potential energy in matter through the neutrino coupling with 
the axial--vector currents of the matter constituents [18--26]. 
For a very lucid and detailed discussion of matter polarisation 
effects on neutrino propagation in a medium we refer the reader to 
ref. \cite{NSSV}; here we summarize some results that will be relevant 
for our discussion and elaborate on some of them. 
  
The contribution of the polarisation of the medium  to the potential energy 
of a test neutrino $\nu_j$ is of the order $\delta V_i(\nu_j) \sim 
(G_F/\sqrt{2}) g_A^i \langle \lambda_i\rangle_\parallel N_i$. Here $G_F$ is 
the Fermi constant, $\langle \lambda_i\rangle_\parallel$, $g_A^i$ and 
$N_i$ are the polarisation along the test neutrino momentum, weak 
axial--vector coupling constant and number density of the particles of the 
$i$th type ($i=e$, $p$, $n$ or background neutrinos). 
Under supernova conditions the average spins of electrons, protons and 
neutrons are rather small; this means that their polarisations are linear in 
the magnetic field strength $B$, i.e. $\delta V_i = c^i B_\parallel$, where 
$B_\parallel$ is the component of the magnetic field along the neutrino 
momentum. 
The electron spin polarisation affects the potential of electron neutrinos 
and antineutrinos in matter through both charged--current and neutral--current 
interactions, whereas for $\nu_\mu$, $\nu_\tau$ and their antineutrinos there 
are only neutral--current contributions $c_Z^e$.  
The polarisations of protons and neutrons contribute to the neutrino 
potentials in a medium only through the neutral--current interactions; 
therefore these contributions are the same for neutrinos of all flavours.  

Thus, one arrives at the following expressions for the neutrino potentials 
in a magnetised medium: 
\be
V(\nu_e)=-V({\bar\nu}_e)=\sqrt{2}G_F(N_e-N_n/2+2N_{\nu_e})+(c_W^e+c_Z^e+c^p
+c^n)B_\parallel\,,
\label{Ve}
\ee
\be
V(\nu_{\mu,\tau})=-V(\bar{\nu}_{\mu,\tau})=\sqrt{2}G_F(-N_n/2 +N_{\nu_e})+
(c_Z^e+c^p+c^n)B_\parallel\,.
\label{Vmu}
\ee
Here $N_e\equiv N_{e^-}-N_{e^+}$, etc., and we have taken into account that 
the number of muon and tauon neutrinos coincides with the number of their 
antineutrinos in supernovae. The electron neutrino number density is 
relatively small in the regions of interest to us in supernovae, and from 
now on we neglect it. Taking it into account would not change our estimates 
significantly.     

The contributions of the electron spin polarisation to the neutrino 
potentials $V(\nu_i)$ were calculated in [18--26]. 
The charged--current electron polarisation contribution to $V(\nu_e)$, 
which we denoted by $c_W^e$, turns out to be twice as large as the 
neutral--current one, and has the opposite sign; as a result, 
\be
c^e = c_Z^e+c_W^e=c_Z^e-2c_Z^e=-c_Z^e\,.
\label{c}
\ee
During the supernova explosion neutrinos and anitineutrinos of all flavours 
are thermally produced in the hot central part of the star. They are  
trapped in the dense core of the supernova and diffuse out on a time 
scale of $\sim 10$ s. When they reach the regions with density $\rho\sim 
(10^{11}-10^{12})\,g/cm^3$, they are no longer trapped and escape freely 
from the star. The surface at unit optical depth is called the 
neutrinosphere. It is located deeper inside the star for $\nu_\mu$, 
$\nu_\tau$ and their antiparticles than for $\nu_e$ and $\bar{\nu}_e$ 
since the medium is more opaque for these latter particles, which have 
charged--current interactions as well as neutral--current ones. As a 
result, the non-electron type neutrinos are emitted at a higher 
temperature, about 6 MeV as against 3 MeV for $\nu_e$ and slightly higher 
for $\bar{\nu}_e$ (because the medium contains more neutrons than 
protons). 

In the supernova environment in the vicinity of the neutrinosphere 
electrons are relativistic and degenerate. In this case \cite{Dolivo,Esp,Elm}
\be 
c_Z^e\simeq \frac{eG_F}{2\sqrt{2}}\left(\frac{3N_e}{\pi^4}\right)^{1/3}\,. 
\label{ce}
\ee
The effects of the polarisation of protons have been calculated in \cite{Esp} 
in the approximation of Dirac protons, i.e. treating protons essentially 
as positrons (though with a different mass). This is certainly not a valid 
approximation since the anomalous magnetic moment of the proton, which is 
neglected in the Dirac approximation, is even larger than its normal 
magnetic moment. Also, the strong--interaction renormalisation of the proton 
axial--vector coupling constant was not taken into account. These 
shortcomings, however, can be readily removed. As we shall see, the nucleons 
are non-relativistic and non-degenerate in the hot proto-neutron star during 
the thermal neutrino emission stage. It is not difficult to calculate their 
polarisations directly using the well-known expressions for the Hamiltonian 
of a non-relativistic fermion in a magnetic field and for the Boltzmann 
distribution function. This gives 
\be
c^p\simeq \frac{G_F}{\sqrt{2}}\, g_A \frac{\mu_p \mu_N}{T}N_p\,,\quad\quad
\quad c^n\simeq -\frac{G_F}{\sqrt{2}}\, g_A \frac{\mu_n \mu_N}{T}N_n\,. 
\label{cpcn}
\ee
Here $\mu_p=2.793$ and $\mu_n=-1.913$ are the proton and neutron magnetic 
moments in units of the nuclear magneton $\mu_N=e/2m_p=3.152\times 10^{-18}$ 
MeV/G, and $g_A=1.26$ is the axial--vector renormalisation constant of the 
nucleon. The expressions in eq. (\ref{cpcn}) coincide with the results 
obtained in the recent paper \cite{NSSV}. 
Notice that $c^p$, $c^n$ and $c_Z^e$ all have the same sign. 
For non-degenerate particles, thermal fluctuations tend to destroy the 
polarisation, hence $c^p$ and $c^n$ decrease with increasing temperature 
$T$. The polarisation of the protons and neutrons in the medium can influence 
the oscillations between the active and sterile neutrinos as well as the 
neutrino spin and spin--flavour precession. However, these effects have not 
been taken into account in most of the previous analyses of neutrino 
conversions in supernovae (the only exception we are aware 
of is ref. \cite{NSSV}). Probably, the reason for this was the general 
idea that heavy nucleons will be polarised to a lesser extent than the   
light electrons. We shall show now that this is not quite correct: even 
though the nucleon polarisation contributions to neutrino potentials are 
typically smaller than that of the polarised electrons in the supernova 
environment, they are of the same order of magnitude. More importantly, 
as we shall see, for spin--flavour precession 
mediated by neutrino transition magnetic moments of Majorana type, the 
effects of polarised electrons nearly cancel out, and nucleon polarisation 
constitutes the main magnetic field effect on the resonance conditions. 

Let us consider now the degree of degeneracy of nucleons in the 
proto-neutron star. The degeneracy parameter of non-relativistic nucleons 
can be written as 
\be
\frac{|\kappa_i|}{T}=\ln\left[\frac{2}{N_i}\left(\frac{m_N T}
{2\pi}\right)^{3/2} \right]\approx \ln\left[41.5\;Y_i^{-1}
\left(\frac{10^{11}\,g/cm^3}{\rho}\right)\left(\frac{T}{3\,{\rm MeV}}
\right)^{3/2}\right]\,.
\label{deg}
\ee
Here $m_N$ is the nucleon mass, $N_i$ and $\kappa_i$ $(i=p,\; n)$ are the 
nucleon number densities and chemical potentials. 
When $d_i\equiv \exp{(\kappa_i/T)}=\exp{(-|\kappa_i|/T)} \ll 1$ the 
nucleons are non-degenerate, i.e. form a classical gas, whereas in the 
opposite limit $d_i\gg 1$ the nucleons will be strongly degenerate.  
In the vicinity of the neutrinosphere ($\rho\sim (10^{11}-10^{12})
\;g/cm^3$, $T\sim (3-6)$ MeV, $Y_e\sim 0.1-0.2$) we have $d_n\aprle 0.08$, 
$d_p\aprle 0.02$, 
i.e. the nucleons are strongly non-degenerate. Therefore the formulas of 
eq. (\ref{cpcn}) are valid there. In the core of the star the densities are 
$\rho\aprge 10^{14}\;g/cm^3$, and the temperatures are higher, too: 
$T\aprge 20$ MeV \cite{Burr1}. As a result, nucleons are weakly 
non-degenerate there with $d_i\sim 1$. This means that one can use 
eq. (\ref{cpcn}) only for rough estimates of the nucleon polarisation 
contribution to neutrino dispersion relations in the core of the supernova.

It is instructive to estimate the relative size of the nucleon and 
electron polarisation contributions to the neutrino potentials. 
Let $Y_i$ be the number of particles of the $i$th kind per baryon. 
Then $N_i=Y_i N$, where $N$ is the total baryon 
number density. The electric neutrality of matter implies that $Y_p=Y_e$ 
(we neglect the nuclei since their fraction is very small in the region of 
interest to us). From eqs. (\ref{ce}) and (\ref{cpcn}) we get 
\be
\frac{c^i}{c_Z^e}=g_A\,\mu_i\,
\left(\frac{\pi}{6}\right)^{1/3}\,\left(\frac{Y_i}{Y_e}\right)^{1/3} 
\left[\frac{N_i}{2}\left(\frac{2\pi}{m_N T}\right)^{3/2} \right]^{2/3}
=g_A\,\mu_i\,\left(\frac{\pi}{6}\right)^{1/3}\,\left(\frac{Y_i}{Y_e}
\right)^{1/3}d_i^{2/3}\,.
\label{cice}
\ee
With increasing degeneracy, the relative contribution of the nucleon 
polarisation increases. However, even in the non-degenerate 
case the nucleon polarisation contributions may not be small provided the 
degeneracy parameter is not too small. 
To see this, let us rewrite $c_p/c_e$ and $c_n/c_e$ in the following form:
\be
\frac{c^p}{c_Z^e}\approx 0.24\left ( \frac{\rho}{10^{11}\,g\,
cm^{-3}}Y_e \right )^{2/3}\left(\frac{ 3\;{\rm MeV}}{T}\right )\,,
\label{cpce}
\ee
\be
\frac{c^n}{c_Z^e}
\approx 0.16 \left ( \frac{\rho}{10^{11}\,g\,cm^{-3}} \right )^{2/3}
\frac{Y_n}{Y_e^{1/3}}\left(\frac{ 3\;{\rm MeV}}{T}\right )\,.
\label{cnce}
\ee
Taking for an estimate $Y_e\approx 0.2$, $Y_n\approx 0.8$, $\rho\approx 
10^{11} \,g/cm^3$ and $T\approx 3$ MeV, we obtain $(c^p+c^n)/c^e\approx 
0.3$, i.e. the nucleon polarisation effect is about 30\% of the electron 
polarisation one. The relative contribution of neutrons and protons is 
$c^n/c^p\approx 0.68(Y_n/Y_e)$; since $Y_n\gg Y_e$ in the supernova, $c^n$ 
dominates over $c^p$. Similar conclusions have been reached in \cite{NSSV}. 

\section{Resonant spin--flavor precession in matter and magnetic fields}
We shall summarize here the main features of the resonant spin--flavour 
precession (RSFP) of neutrinos in matter and a magnetic field 
\cite{Akhm1,LM} and compare them with the corresponding characteristics of 
the MSW effect \cite{MSW}; for a more detailed discussion of the RSFP see 
\cite{Akhm2}. 

Neutrinos with non-vanishing flavour-off-diagonal (transition) magnetic 
moments experience a simultaneous rotation of their spin and flavour in a 
transverse magnetic field (spin--flavour precession) \cite{ScV}. 
In vacuum, such a precession is suppressed because of the kinetic energy 
difference of neutrinos of 
different flavours, $\Delta E_{kin}\simeq \Delta m^2/2E$ for relativistic 
neutrinos. At the same time, in matter this kinetic energy difference can 
be cancelled by the potential energy difference $V(\nu_i)-V(\bar{\nu}_j)$, 
leading to a resonant enhancement of the spin--flavour precession 
\cite{Akhm1,LM}. 
The RSFP of neutrinos is similar to resonant neutrino oscillations 
(MSW effect \cite{MSW}). For Dirac neutrinos, 
their transition magnetic moments cause transitions between left--handed 
neutrinos of a given flavour and right--handed (sterile)
neutrinos of a different flavour. For Majorana neutrinos the spin--flavour 
precession due to their transition magnetic moments induces transitions 
between left--handed neutrinos of a given flavour and right--handed 
antineutrinos of a different flavour which are not sterile. 

\subsection{Resonance conditions}
The resonance condition for a transition between left--handed 
neutrinos of the $i$th flavour and right--handed neutrinos or antineutrinos  
of the $j$th flavour ($i,j=e\,\mu\,,\tau$ or $s$ where $s$ means the sterile
neutrino) is 
\be
V(\nu_{iL})+\frac{m_{\nu_i}^2}{2E}=V(\bar{\nu}_{jR})+\frac{m_{\nu_j}^2}{2E}\,. 
\label{resij}
\ee
For antineutrinos of the $i$th type $V(\bar{\nu}_{iR})=-V(\nu_{iL})$. Mean 
potential energies of active neutrinos and antineutrinos 
including matter polarisation effects are given in eqs. (\ref{Ve}) and 
(\ref{Vmu}); for sterile neutrinos $V(\nu_s)=0$. 

The resonance conditions for various RSFP transitions can be written in 
the following generic form:
\be
\sqrt{2}G_F N_{\rm eff}-c_{\rm eff}B_\parallel=
\frac{\Delta m^2}{2E}\,.
\label{res}
\ee
The resonance condition for neutrino oscillations in matter (MSW effect) 
has almost the same form, the difference being that the r.h.s. of 
eq. (\ref{res}) is multiplied by the cosine of the double vacuum mixing 
angle, $\cos 2\theta_0$. The effective parameters of the resonance 
condition depend on the nature of the transition in question. In the 
following table we summarize the parameters that enter into the resonance 
condition (\ref{res}) for the neutrino conversions of interest to us.

\vspace{0.2cm}
Table 1.

\begin{tabular}{|c||c|c|c|c||} \hline
No & transition & $N_{\rm eff}$ & $c_{\rm eff}$ & $\Delta m^2$ \\
\hline
1 & $\nu_{e}\leftrightarrow \bar{\nu}_{x}$ & $N_n-N_e$ 
& $2(c^p+c^n)$ & $m_{\nu_{e}}^2-m_{\nu_{x}}^2$ \\ 
2 & $\bar{\nu}_{e}\leftrightarrow \nu_{x}$ & $N_n-N_e$ 
& $2(c^p+c^n)$ & $m_{\nu_{x}}^2-m_{\nu_{e}}^2$ \\ 
3 & $\nu_{e}\leftrightarrow \bar{\nu}_{s}$ & $N_e-N_n/2$ &
$c_Z^e-c^p-c^n$ & $m_{\nu_{s}}^2-m_{\nu_{e}}^2$ \\
4 & $\bar{\nu}_{e}\leftrightarrow \nu_{s}$ & $N_e-N_n/2$ &
$c_Z^e-c^p-c^n$ & $m_{\nu_{e}}^2-m_{\nu_{s}}^2$ \\
5 & $\nu_{x}\leftrightarrow \bar{\nu}_{s}$ & $N_n/2$ & 
$c_Z^e+c^p+c^n$ & $m_{\nu_{x}}^2-m_{\nu_{s}}^2$ \\
6 & $\bar{\nu}_{x}\leftrightarrow \nu_{s}$ & $N_n/2$ & 
$c_Z^e+c^p+c^n$ & $m_{\nu_{s}}^2-m_{\nu_{x}}^2$ \\
\hline
7 & $\nu_{e}\leftrightarrow \nu_{x}$ & $N_e$ 
& $2 c_Z^e$ & $m_{\nu_{x}}^2-m_{\nu_{e}}^2$ \\ 
8 & $\bar{\nu}_{e}\leftrightarrow \bar{\nu}_{x}$ & $N_e$ 
& $2 c_Z^e$ & $m_{\nu_{e}}^2-m_{\nu_{x}}^2$ \\ 
9 & $\nu_{e}\leftrightarrow \nu_{s}$ & $N_e-N_n/2$ &
$c_Z^e-c^p-c^n$ & $m_{\nu_{s}}^2-m_{\nu_{e}}^2$ \\
10 & $\bar{\nu}_{e}\leftrightarrow \bar{\nu}_{s}$ & $N_e-N_n/2$ &
$c_Z^e-c^p-c^n$ & $m_{\nu_{e}}^2-m_{\nu_{s}}^2$ \\
11 & $\nu_{x}\leftrightarrow \nu_{s}$ & $N_n/2$ & 
$c_Z^e+c^p+c^n$ & $m_{\nu_{x}}^2-m_{\nu_{s}}^2$ \\
12 & $\bar{\nu}_{x}\leftrightarrow \bar{\nu}_{s}$ & $N_n/2$ & 
$c_Z^e+c^p+c^n$ & $m_{\nu_{s}}^2-m_{\nu_{x}}^2$ \\
\hline
\end{tabular}

\vspace{0.2cm}
\noindent
Here $\nu_s$ is a sterile neutrino which is assumed to be left--handed 
($\bar{\nu}_s$ is right--handed),  
$\nu_x=\nu_\mu$ or $\nu_\tau$, and for the sake of comparison we have 
also included the parameters for the neutrino conversions due to the 
MSW effect (lines 7--12). For $c_{\rm eff}B_\parallel<\sqrt{2}G_F 
N_{\rm eff}$, the neutrino transitions listed in Table 1 can 
only be resonantly enhanced if the corresponding $\Delta m^2$ 
and $N_{\rm eff}$ are of the same sign. 
For given signs of $N_{\rm eff}$, only 6 of the 12 transitions can be 
resonant, depending on the signs of the respective $\Delta m^2$.  
We shall comment on the $c_{\rm eff}B_\parallel>\sqrt{2}G_F N_{\rm eff}$  
case later on. It is interesting to notice that 
the parameters of the resonance conditions for the RSFP transitions 
involving sterile neutrinos or antineutrinos and those for the corresponding 
MSW transitions are the same (lines 3--6 and 9--12 respectively). 
The reason for this is that $V(\nu_s)=0=V(\bar{\nu}_s)$.

Several remarks are in order. The authors of \cite{SSV} pointed out 
that the electron polarisation contribution cancels out in the  
resonance condition for the RSFP transitions $\nu_{e}\leftrightarrow 
\bar{\nu}_{x}$ and $\bar{\nu}_{e}\leftrightarrow \nu_{x}$. They therefore
concluded that magnetic fields have no effect on these 
resonance conditions. We would like to emphasize that the cancellation 
noticed in \cite{SSV} is not exact; it holds only up to the electroweak 
radiative corrections. In particular, this cancellation is a consequence 
of eq. (\ref{c}) which is based on the tree-level relation between the 
masses of the $W^\pm$ and $Z^0$ gauge bosons, $M_W^2=M_Z^2 \cos^2\theta_W$. 
This relation is known to receive radiative corrections of the order 
of $0.5\%$ (mainly due to the heavy top quark). There 
are other electroweak corrections that would also lead to an incomplete 
cancellation, e.g. from the anomalous magnetic moment of the electron.
They are typically of the same order of magnitude, $\aprle 0.5\%$. However, 
for supernovae more important contributions come from the polarisation of 
nucleons in the magnetised medium. As was demonstrated in sec. 2, 
they are quite sizable and can be comparable with $c^e$.

In ref. \cite{KS3} it has been claimed that the potential of electron 
neutrinos $V(\nu_e)$ does not receive any matter-polarisation contributions. 
The authors, following ref. \cite{Esp}, claimed that the electron and proton 
polarisation effects cancel each other in $V(\nu_e)$ and therefore concluded 
that the $\nu_e\leftrightarrow \nu_s$ and $\bar{\nu}_e\leftrightarrow 
\bar{\nu}_s$ oscillations in the supernova are not affected by the magnetic 
field and so cannot be the cause of the pulsar birth velocities. The 
cancellation was the result of the assumption that protons are strongly 
degenerate in the proto-neutron star; as we have shown in sec. 2, this 
assumption is incorrect. Moreover, even for degenerate protons the 
cancellation takes place only if they are treated as Dirac particles, 
i.e. when the strong--interaction renormalisation of the proton's magnetic 
moment and axial--vector coupling constant are neglected.  
In addition, the effects of neutron polarisation, which can be comparable 
with those of polarised electrons, were not considered in \cite{KS3}. It 
should be noted, however, that the above shortcomings have no effect on the 
$\nu_{x}\leftrightarrow \nu_s$ transitions which were the main topic 
of ref. \cite{KS3}. 

In refs. \cite{DN,SSV} it was claimed that in strong enough magnetic 
fields the term $c^e B_{\parallel}$ can overcome the $\sqrt{2}G_F N_e$ 
term in the resonance condition of the $\nu_e \leftrightarrow \nu_{x}$ and 
$\bar{\nu}_e \leftrightarrow \bar{\nu}_{x}$ oscillations, leading to the 
possibility of new resonances. However, it has been demonstrated in 
\cite{NSSV} that this is incorrect: the electron polarisation contribution 
can never exceed the electron density contribution, 
essentially because the mean polarisation of electrons cannot exceed 
unity. The expressions for the polarisation of matter constituents 
(\ref{ce}) and (\ref{cpcn}) that are linear in the magnetic field are  valid 
only when the induced polarisations of the particles in matter are small.
Nevertheless, as was stressed in \cite{NSSV}, for the $\nu_e \leftrightarrow 
\nu_{s}$ and $\bar{\nu}_e \leftrightarrow \bar{\nu}_{s}$ oscillations 
the $c_{\rm eff} B_{\parallel}$ term can indeed overcome the $\sqrt{2}G_F 
N_{\rm eff}$ term in the resonance condition, since the effective number 
density $N_{\rm eff}=N_e-N_n/2$ can become small provided that there is a 
compensation between the electron and neutron contributions. In this case 
new resonances are indeed possible, and the resonant channel will depend 
on whether the neutrinos are emitted along the magnetic field or in the 
opposite direction. Obviously, the same argument applies for the RSFP 
transitions involving sterile neutrinos or antineutrinos (lines 3--6 of 
Table 1). We would like to point out here that in the case of the RSFP 
transitions the same is also true even for the  $\nu_e \leftrightarrow 
\bar{\nu}_{x}$ and $\bar{\nu}_e \leftrightarrow \nu_{x}$ oscillations 
(lines 1 and 2 of Table 1) that involve only active neutrinos; the matter 
polarisation term can overcome the $\sqrt{2}G_F N_{\rm eff}$ term provided 
that the effective density $N_{\rm eff}=N_n-N_e$ becomes small because of 
the compensation of the electron and neutron number densities. 
Thus, in the case of the RSFP of neutrinos, the matter polarisation term 
in the resonance condition can in principle exceed the effective matter 
density term leading to the possibility of new resonances {\em for all 
types of neutrino transitions}.

\subsection{Adiabaticity parameters and transition probabilities}
The probability of an RSFP--induced neutrino transition depends on the 
degree of its adiabaticity and is generally large when the adiabaticity 
parameter $\gamma$ is large enough:   
\be
\gamma=4\,\frac{(\mu_\nu B_{\bot r})^{2}}{\sqrt{2}G_F N_{\rm eff}}\, 
L_{\rho r}\approx 0.81\, Y_{\rm eff}^{-1}\,\left(\frac{10^{11}\,g/cm^3}
{\rho}\right) \left[\frac{\mu_{\nu}}{10^{-13}\mu_B}\;\frac{B_{\bot 
r}}{3\times 10^{14}\,{\rm G}}\right]^2\left(\frac{L_{\rho r}}{10\,{\rm km}}
\right)>1\,. 
\label{adiab2}
\ee
Here $\mu_{\nu}$ is the neutrino transition magnetic moment, $B_{\bot r}$ is 
the strength of the transverse component of the magnetic field at the 
resonance, $Y_{\rm eff}$ is defined through $N_{\rm eff}=Y_{\rm eff}N$, and 
$L_{\rho}\equiv \vert \frac {1}{\rho}\frac{d\rho}{dr} \vert^{-1}$ is the 
characteristic length over which the matter density varies significantly in 
the supernova, $L_{\rho r}$ being its value at the resonance. 

The spin--flavour precession 
is typically strongly  suppressed far below and far above the resonance 
point; in this case the transition probability is to a very good accuracy 
approximated by 
\be
P_{tr}\approx (1-P')\,,\quad\quad P'\equiv \exp{(-\frac{\pi}{2}
\gamma)}\,.
\label{Ptr}
\ee
The probability of neutrino transitions due to the MSW effect is given 
by the same expression with the adiabaticity 
parameter $\gamma$ replaced by $\gamma_{\rm MSW}=(\sin^2 2\theta_0/
\cos 2\theta_0)(\Delta m^2/2E)L_{\rho r}$.  
\footnote{We note in passing that the MSW adiabaticity condition was 
formulated incorrectly in refs. \cite{KS1} and \cite{KS3}: The 
oscillation length at resonance must be compared with the resonance 
width $\Delta r=2\tan 2\theta_0\,L_{\rho r}$ and not with $L_{\rho r}$ 
itself.} For adiabatic transitions ($\gamma \gg 1$) the transition 
probability is close to one. 
In the vicinity of the neutrinosphere, $L_{\rho r}\sim 10$ km; 
therefore the RSFP transitions will be adiabatic for 
\be
B_{\bot r}\aprge 3\times 10^{14}\,(10^{-13}\mu_B/\mu_\nu)\;{\rm G}\,. 
\label{constr1}
\ee
This constraint can in principle be somewhat relaxed since $Y_{\rm eff}$ is 
typically $<1$ and can also be $\ll 1$ in some cases; hovewer, we 
will need the RSFP adiabaticity condition to be satisfied with some 
margin and so shall continue to use eq. (\ref{constr1}). For the MSW 
transitions to 
be adiabatic near the neutrinosphere the vacuum mixing angle should satisfy 
$\theta_0\aprge 10^{-4}$.

\section{Kick momenta of pulsars}
In ref. \cite{KS1} the following two conditions for generating the 
pulsar kick velocity through resonant neutrino oscillations were 
formulated: 

(1) The neutrino conversion takes place between the neutrinospheres of 
two different neutrino species;

(2) The resonance coordinate depends on the angle between the directions of 
the magnetic field and the neutrino momentum. 

\noindent
These conditions apply to the RSFP--induced neutrino conversions as well. 
Consider, e.g., the $\nu_\tau \leftrightarrow \bar{\nu}_e$ conversions above 
the $\nu_\tau$--sphere but below the $\bar{\nu}_e$--sphere. A $\nu_\tau$ 
propagates freely until it gets transformed into $\bar{\nu}_e$  through the 
RSFP conversion. The resulting $\bar{\nu}_e$ cannot escape easily and gets 
trapped since the resonance point is below its neutrinosphere. At the 
same time, a $\bar{\nu}_e$ which initially was trapped and diffused out 
slowly will be converted into $\nu_\tau$ when it reaches the resonance 
surface. The resulting $\nu_\tau$ escapes freely since the resonant 
conversion occurred above its neutrinosphere. Thus, the resonance surface 
becomes the new ``neutrinosphere'' for the $\nu_\tau$'s. It is, however, not 
a sphere. The matter polarisation in the supernova magnetic field leads to 
the resonance taking place at different distances from the core of the star 
for neutrinos emitted parallel and antiparallel to the magnetic field; as a 
result, the resonance surface has different temperatures in these two 
directions leading to an asymmetry of the momenta of the emitted neutrinos. 

Let us  estimate the magnitudes of $\Delta m^2$ that are 
necessary for various neutrino conversions to occur in the regions of 
interest to us.  Since the neutrino mean energy is 
$\langle E\rangle \approx 3.15 T$, from eq. (\ref{res}) one finds 
\be
\Delta m^2\approx 1.4 \times 10^5 \, Y_{\rm eff}\left(\frac{\rho}{10^{11}\,
g/cm^3}\right)\left(\frac{T}{3\, {\rm MeV}}\right)\,.
\label{deltam2}
\ee
This depends on $Y_{\rm eff}$ and therefore on the neutrino transition in 
question. In the vicinity of the neutrinosphere the electron fraction $Y_e$ 
is typically of the order of 0.1--0.2 at the time when neutrinos are 
copiously produced in the supernova (a few seconds after the core bounce).
It decreases towards smaller densities and reaches the value of about 0.46. 
At the same time, in the dense core of the supernova there is still a 
significant amount of trapped $\nu_e$'s which hinder the neutronization 
process. Therefore $Y_e$ can be close to 0.4 in the supernova's core 
(see \cite{BO} for a more detailed discussion). 

For our estimates we will assume a hierarchical pattern of neutrino masses. 
For MSW transitions between active neutrinos or antineutrinos (lines 
7 and 8 of Table 1) the required heavier neutrino mass 
is in general in the range $m_2 \sim (100-800)$ eV. 
For the RSFP transitions between active neutrinos and antineutrinos 
one would need $m_2\sim (300-1500)$ eV, whereas for transitions between 
muon or tauon neutrinos or antineutrinos and sterile neutrino states 
(lines 5, 6, 11 and 12 of Table 1) $m_2$ should be in the range $200 \aprle 
m_2 \aprle 1.6\times 10^4$ eV. In all these cases neutrinos do not satisfy 
the cosmological bounds on the mass of stable neutrinos and would have to 
decay sufficiently fast. 
However, for the RSFP and MSW transitions between electron neutrinos or 
antineutrinos and sterile neutrino states (lines 3, 4, 9 and 10 of Table 1) 
$m_2$ can be considerably smaller since $Y_{\rm eff}$ can be very small 
in this case. Indeed, the parameter $Y_e$ passes through the value $1/3$  
somewhere between the supernova core and the neutrinosphere, i.e.  
$Y_{\rm eff}$ passes through zero. This means  that the required value of 
$\Delta m^2\approx m_2^2$ can be very small, too: $0\le m_2 \aprle 10$ keV. 
For example, $m_2$ can well be in the ranges of a few eV or few tens of eV, 
which are both cosmologically safe and interesting (in particular, the 
allowed ranges of neutrino mass and transition magnetic moment are consistent 
with the predictions of the decaying neutrino theory of the 
ionisation of the interstellar medium \cite{Sciama}). 
The mass $m_2$ can also be in the range $10^{-3}-10^{-4}$ eV which 
of interest for the solar neutrino problem; 
moreover, in this case the transition can be resonant even for massless 
neutrinos. For $m_2\aprle 4$ keV transitions including both electron 
neutrinos and antineutrinos can be resonantly enhanced\footnote{This 
includes the case $\Delta m^2=0$ which corresponds to the 
resonance transition due to the ordinary (flavour--diagonal) magnetic 
moments of neutrinos \cite{Vol,Akhm1}).}; 
this would lead to an additional factor of two increase of the pulsar 
velocities for a given value of the supernova magnetic field. 

We will now derive the generalized expression for the 
neutrino momentum asymmetry which will be valid for the RSFP as well as  
for neutrino oscillation transitions. 
The relative recoil momentum of a pulsar can be estimated as \cite{KS1} 
\be
\frac{\Delta k}{k}\propto \frac{T^4(r_0-\delta)-T^4(r_0+\delta)}{T^4(r_0)}\,,
\label{asymm1}
\ee
where $r_0$ is the position of the resonance in the absence of the 
magnetic field and $\pm \delta$ is the shift of the resonance coordinate 
for the neutrinos emitted parallel and antiparallel to the magnetic field. 
{}From the generic resonance condition (\ref{res}) one can estimate the 
value of $\delta$ as 
\be
\delta\simeq\frac{c_{\rm eff}B}{\sqrt{2}G_F (dN_{\rm eff}/dr)}\,,
\label{delta}
\ee
where $B$ is the magnetic field strength. For the neutrino momentum 
asymmetry we get 
\be
\frac{\Delta k}{k}\approx \frac{1}{6}\cdot 2\cdot 4\cdot R\, \frac{1}{T}
\frac{dT}{dr}\,\delta \approx
\frac{4}{3}\,R\, \left(\frac{c_{\rm eff}B}{\sqrt{2}G_F}\right)\frac{1}{T}
\frac{dT}{dN_{\rm eff}}\,.
\label{asymm2}
\ee
Here the factor 1/6 takes into acount that only one neutrino or 
antineutrino species out of 6 acquires a momentum asymmetry, and  $R$ is 
a geometrical factor 
to be discussed below. 
The desirable value of $\Delta k/k$ is about $10^{-2}$; this can be 
achieved for a temperature asymmetry of the order of $10^{-2}$ (we are 
assuming here the neutrino transition probability $P_{tr}\approx 1$). 

For resonant neutrino oscillations, the geometrical factor $R$ in 
eq. (\ref{asymm2}) takes into account the fact that for the neutrinos 
emitted in directions orthogonal to the supernova magnetic field 
the resonance condition is not affected by the field. Therefore such 
neutrinos do not contribute to the kick velocity of the pulsar. 
In order to find this velocity one has to calculate the net  
momentum of neutrinos emitted in the direction of the magnetic field.  
KS estimated the resulting geometrical factor as $1/2$; however, their result 
was criticised by Qian \cite{Qian} who showed that in fact $R \approx 1/6$.  

For spin--flavour precession the situation is somewhat different. 
The magnetic field plays a dual role in this process: first, its component 
$B_\bot$ transverse to the neutrino momentum mixes the left--handed 
and right--handed neutrino states and causes the spin--flavour 
precession itself; second, the longitudinal component $B_\parallel$ 
affects the resonance condition, as discussed in sec. 3.1. Both roles 
are important for the purposes of our discussion. In fact, only the 
neutrinos emitted in directions different from that of the magnetic 
field or orthogonal to it can contribute to $\Delta k/k$. For 
neutrinos emitted exactly along the magnetic field, the RSFP does not 
occur; for those emitted in the orthogonal plane the field has no  
effect on the resonance condition and therefore does not lead to any 
momentum asymmetry. For this reason, in the case of the RSFP, the 
geometrical factor in eq. (\ref{asymm2}) can be written as $R=R_1 R_2$,  
where $R_1\approx 1/6$ as for neutrino oscillations, while 
the factor $R_2$ takes into account the reduction of the RSFP 
transition probability $P_{tr}$ for neutrinos emitted close to the magnetic 
field directions. Basically, this reduction excludes the zenith angles 
close to $0$ and $180^\circ$ in the angular integration over the neutrino 
momenta. The numerical value of $R_2$ depends on the extent of 
the excluded region, which in turn depends on the magnitude of the 
adiabaticity parameter $\gamma$. If the adiabaticity condition 
(\ref{adiab2}) is satisfied with a large margin (i.e. $\gamma \gg 1$), 
even a strong reduction of $B_\bot=B\sin\theta$ because of the zenith angle 
$\theta$ being close to 0 or $180^\circ$ will not suppress the RSFP 
transition probability significantly. In this case $R_2\approx 1$ and 
$R\approx R_1\approx 1/6$. In our estimates we will assume that this 
is the case.

An important parameter that enters into the neutrino momentum asymmetry 
(\ref{asymm2}) is the derivative $dT/dN_{\rm eff}=(dN_{\rm eff}/dT)^{-1}$.
The effective density $N_{\rm eff}$ is in general a linear combination of 
$N_e$ and $N_n$, depending on the type of the neutrino conversion. 
The derivative $dN_e/dT$ was estimated by KS as 
$(\partial N_e/\partial T)_{\kappa_e}$ using the relativistic Fermi 
distribution function for the electrons. 
This gave 
\be
dN_e/dT\approx \frac{2}{3}(3\pi^2 N_e)^{1/3} T\,.
\label{dNedTwrong}
\ee
However, this approach was criticised by 
Qian \cite{Qian} who pointed out that the chemical potential of electrons 
cannot be considered as temperature independent in the supernova. He 
suggested to use instead the results of numerical simulations of matter 
density and temperature profiles, which typically give 
$N \propto T^3$ \cite{BM}. We adopt this approach here. 

Using $N_i=Y_i N$ ($i=e,n$) it is easy to show that    
\be
\frac{dN_i}{dT}\approx \frac{N_i}{T}\left( 3+\frac{d\ln Y_i}{d\ln T}
\right) \,.
\label{dNidT}
\ee
The electron fraction $Y_e$ decreases (and therefore $Y_n$ increases) with 
increasing $r$ below the $\nu_e$-sphere. For this reason for electrons the 
expression in the parentheses in eq. (\ref{dNidT}) is in fact larger than 
3 whereas for neutrons it is slightly smaller than 3. 
Estimates of the logarithmic derivatives using the $Y_e$ profile from 
\cite{BL} give 
\be
\frac{dN_e}{dT}\approx 4\frac{N_e}{T}\,,\quad\quad\quad 
\frac{dN_n}{dT}\approx 2.8\frac{N_n}{T}
\label{dNdT}
\ee
in the neutrinospheric region. Notice that numerically $dN_e/dT$ in eq. 
(\ref{dNdT}) is about an order of magnitude larger than the corresponding KS 
value (\ref{dNedTwrong}) \cite{Qian}. 


It is instructive to estimate the relative sizes of $dN_n/dT$ and 
$dN_e/dT$ in the supernova environment: 
\be
\frac{dN_n/dT}{dN_e/dT}\approx 
0.7\, (Y_n/Y_e)\,.
\label{ratio1}
\ee
Thus, $dN_n/dT$ is typically a factor of 3 to 6 larger than $dN_e/dT$. 
%
%
Notice that the $dN_e/dT$ contribution to 
$dN_{\rm eff}/dT$ has the opposite sign compared to the $dN_n/dT$ one 
(lines 1--4, 9 and 10 of Table 1) and so will tend to increase the kick, 
especially for transitions of electron neutrinos and antineutrinos 
into sterile states. 

We shall first estimate the asymmetry $\Delta k/k$ for the RSFP--induced 
transitions between active neutrinos $\nu_e\leftrightarrow \bar{\nu}_x$ and 
$\bar{\nu}_e\leftrightarrow \nu_x$ due to the Majorana neutrino transition 
magnetic moments (lines 1 and 2 of Table 1): 
\be
\frac{\Delta k}{k}\approx \frac{2}{9} \left[\frac{2 (c^p+c^n)B}{\sqrt{2}G_F}
\right]\frac{1}{T} \frac{dT}{d(N_n-N_e)} \approx 1.2\times 10^{-4}
\left(\frac{B}{3\times 10^{14}\,{\rm G}}\right) \frac{3\,{\rm MeV}}{T}\,. 
\label{asymm3}
\ee
Let us compare eq. (\ref{asymm3}) with the corresponding expression for 
the case of the $\nu_e\leftrightarrow \nu_x$ oscillations derived in  
\cite{KS1}. 
We first notice that the numerical coefficient in eq. (10) of \cite{KS1} 
was overestimated (and so the requisite magnetic field underestimated) by 
about a factor of 40, where a factor $\sim 3$ comes from the geometrical 
factor $R$ and a factor $\sim 13$ from $dN_e/dT$ \cite{Qian}.  
Apart from the different numerical coefficient, their expression for 
$\Delta k/k$ falls with increasing temperature as $T^{-2}$ and not as 
$T^{-1}$. Comparing eq. (\ref{asymm3}) with 
the corrected 
eq. (10) of ref. \cite{KS1} 
we find that in order to obtain the same effect on the pulsar velocities one 
would need about a factor of two stronger magnetic field in the case of the 
RSFP of active neutrinos than in the case of the resonant oscillation of 
active neutrinos. For example, in order to produce $\Delta k/k\simeq 1\%$ a 
magnetic field $B\aprge 2.5\times 10^{16}$ G is necessary. This field is of 
the same order of magnitude as that needed to explain the pulsar birth 
velocities by asymmetric neutrino production \cite{Field-weak}.
By contrast, in our case, neutrinos carrying an asymmetric momentum will 
experience very few interactions with matter, and so the asymmetry is 
unlikely to be suppressed by such interactions.  

Next, we consider the RSFP--induced transitions between active and 
sterile neutrino states due to the Dirac neutrino transition magnetic 
moments (lines 3-6 of Table 1). For the transitions $\nu_x \leftrightarrow 
\bar{\nu}_s$ and $\bar{\nu}_x \leftrightarrow \nu_s$ we obtain 
\begin{eqnarray}
\frac{\Delta k}{k} & \approx & (2/9) \left\{[c_Z^e+ (c^p+c^n)]\, B/
\sqrt{2}G_F \right\}\,(1/T)\, [dT/d(N_n/2)] \nonumber \\ & \approx &
\left[3.7\times 10^{-4}\, \frac{Y_e^{1/3}}{Y_n} \left( \frac{10^{11}\,g/cm^3}
{\rho}\right)^{2/3}+ 7.6\times 10^{-5}\left(\frac{3\,{\rm MeV}}{T}\right)
\right]\left(\frac{B}{3\times 10^{14}\,{\rm G}}\right). 
\label{asymm4}
\end{eqnarray}
{}From eq. (\ref{asymm4}) it follows that in order to 
get $\Delta k/k\simeq 1\%$ one would need $B\aprge 9\times 10^{15}$ G. 
This field is of the same order of magnitude as the one that is 
needed in the case of the neutrino flavour oscillations. 
Moreover, for a hierarchical neutrino mass pattern $m_{\nu_s} \gg 
m_{\nu_\mu},m_{\nu_\tau}$ the transitions between both $\bar{\nu}_\mu$  
and $\bar{\nu}_\tau$ and sterile neutrino states can be resonant and 
contribute to $\Delta k/k$ 
\footnote{This has been pointed out for the case of oscillations into 
sterile neutrinos in \cite{KS3}.}. 
In this case a factor of two weaker field would be able to 
produce the desired kick. 

For the transitions $\nu_e \leftrightarrow \bar{\nu}_s$ and $\bar{\nu}_e 
\leftrightarrow \nu_s$ we obtain 
$$\frac{\Delta k}{k}  \approx  (2/9) \left\{[c_Z^e- (c^p+c^n)]\, B/
\sqrt{2}G_F \right\}\,(1/T)\, [dT/d(N_n/2-N_e)] $$
\be 
\approx \left[3.7\times 10^{-4}\, \frac{Y_e^{1/3}}{Y_n} 
\left( \frac{10^{11}\,g/cm^3}{\rho}\right)^{2/3}-7.6\times 10^{-5}
\left(\frac{3\,{\rm MeV}}{T}\right)\right]
\frac{Y_n}{Y_n-2.86 Y_e}
\left(\frac{B}{3\times 10^{14}\,{\rm G}}\right). 
\label{asymm5}
\ee
For these transitions in order to get $\Delta k/k\simeq 1\%$ one would 
typically need $B\aprge 4\times 10^{15}$ G. This field is about a factor 
of two weaker than the one that is needed in the case of the KS mechanism. 
We would like to ephasize, however, that our consideration is rather 
simplified and can only yield order-of-magnitude estimate ot the 
requisite magnetic field strengths. 

It should be noticed that the results in eqs. (\ref{asymm4}) and 
(\ref{asymm5}) apply to 
the case of resonant oscillations between active and sterile neutrinos  
as well. This case was studied in \cite{KS3}; however the result 
obtained in that paper differs from ours. The reason for this is that 
the authors of \cite{KS3} erroneously considered neutrons as strongly 
degenerate in the hot proto-neutron star. This resulted in a suppressed 
value of $\Delta k/k$, and in order to save the situation, they had to assume 
that the resonant oscillations take place deep in the core of the supernova. 
However, as follows from our considerations, 
the non-degeneracy of neutrons increases $\Delta k/k$ so that there is no 
need to assume that the resonance takes place in the supernova's core. 
Moreover, the asymmetry decreases with the resonance density.

The lower bounds on the supernova magnetic fields $B$ were derived here 
assuming that the RSFP transitions are adiabatic. The adiabaticity condition 
puts another lower bound on $B$, eq. (\ref{constr1}). The bounds obtained in 
this section are more restrictive provided the neutrino transition magnetic 
moments satisfy $\mu_\nu\aprge 10^{-14}\mu_B$ ($10^{-15}\mu_B$) for Dirac 
(Majorana) neutrinos.

\section{Conclusion}
We have studied the effects of the spin polarisation of matter in a 
supernova magnetic field on the resonance conditions for spin--flavour 
precession of Dirac and Majorana neutrinos in supernovae. 
The magnetic field distorts 
the resonance surface resulting in an asymmetric neutrino emission, which 
can explain the observed space velocities of pulsars and their 
possible correlation with the pulsar magnetic fields. 
Our estimates for the case of spin and spin--flavour precession of Dirac 
neutrinos also apply to  oscillations into sterile neutrinos and correct the 
results of ref. \cite{KS3} where the effect was underestimated.  
In the case of  
resonant spin--flavour precession into sterile neutrino states due to
Dirac transition magnetic moments of neutrinos, the requisite supernova 
magnetic field strengths are $B\aprge 4\times 10^{15}$ G.  
This is about a factor of 2 smaller than the field necessary in the case 
of neutrino flavour oscillations.  
Such fields are considered possible in supernovae \cite{DT}.
For resonant spin--flavour 
precession between active neutrinos and antineutrinos due to Majorana 
transition magnetic moments of neutrinos, magnetic field strengths 
$B\aprge 2\times 10^{16}$ G would be needed. 

The authors are grateful to Alexei Smirnov for useful discussions and to 
the referee for helpful criticism which have led to improvements in our 
arguments. E.A. is grateful to ICTP for financial support. A.L. and D.W.S.  
acknowledge financial support from MURST.

\end{document}